\documentclass[aps,twocolumn,superscriptaddress,showpacs,amsmath,amssymb,footinbib,longbibliography]{revtex4-1}
\usepackage[english]{babel}
\usepackage[utf8]{inputenc}
\usepackage{amssymb,amsbsy,amsmath}
\usepackage{graphicx}
\usepackage{dcolumn}
\usepackage{bm}
\usepackage{subfigure}
\usepackage{braket}
\usepackage[svgnames]{xcolor}
\usepackage{textcomp}
\usepackage[colorlinks,linkcolor=blue,citecolor=blue,urlcolor=cyan]{hyperref}

\newcommand {\peq} {p_{\mathrm{eq.}}}
\newcommand {\tth} {t\,t_{\uparrow}}
\newcommand {\hc} {\mathrm{h.c.}}
\newcommand {\inice} {\mathrm{i}}

\newcommand {\figref}[1]{Fig.~\protect\ref{#1}}

\newcommand {\Tr}[1]{\mbox{Tr}\left[\,{#1}\,\right]}

\newcommand {\emath}[1]{\mathrm{e}^{#1}}
\begin{document}

\title{Density dynamics in the mass-imbalanced Hubbard chain}

\author{Tjark Heitmann}
\email{tjark.heitmann@uos.de}
\author{Jonas Richter}
\affiliation{Fachbereich Physik, Universit{\"a}t
  Osnabr{\"u}ck, Barbarastr. 7, D-49076 Osnabr{\"u}ck, Germany}
\author{Thomas Dahm}
\affiliation{Fakult{\"a}t f{\"u}r Physik, Universit{\"a}t Bielefeld, Postfach 100131, D-33501 Bielefeld, Germany}
\author{Robin Steinigeweg}
\email{rsteinig@uos.de}
\affiliation{Fachbereich Physik, Universit{\"a}t
  Osnabr{\"u}ck, Barbarastr. 7, D-49076 Osnabr{\"u}ck, Germany}

\date{\today}

\begin{abstract}

	We consider two mutually interacting fermionic particle species on a one-dimensional 
	lattice and study how the mass ratio  
	$\eta$ between the two species affects the (equilibration) 
	dynamics of the particles.
	Focussing on the regime of strong interactions and 
	high temperatures, two well-studied points of reference are 
	given by (i) the case of equal masses ${\eta = 1}$, i.e., the standard 
	Fermi-Hubbard chain, where initial non-equilibrium density distributions are 
	known to decay, and (ii) the case 
	of one particle species being infinitely heavy, ${\eta = 0}$, leading to a 
	localization of the lighter particles in an effective disorder potential.
	Given these two opposing cases, the dynamics in the case of intermediate mass ratios ${0 < \eta < 1}$ is of particular interest.
	To this end, we study the real-time dynamics 
	of pure states featuring a sharp initial non-equilibrium 
	density profile. Relying on the concept of dynamical 
	quantum typicality, the resulting non-equilibrium dynamics
	can be related to equilibrium correlation functions.
	Summarizing our main results, we observe that 
	diffusive transport occurs for moderate 
	values of the mass imbalance, and manifests itself in a 
	Gaussian spreading of real-space density profiles and an exponential decay of 
	density modes in momentum space. For stronger 
	imbalances, we provide evidence that transport becomes anomalous on 
	intermediate time scales and, in 
	particular, our results are consistent with the absence of strict 
	localization in the long-time limit for any ${\eta > 0}$. Based on our numerical 
	analysis, we provide an estimate for the ``lifetime'' of the effective 
	localization as a function of $\eta$.

\end{abstract}

 \keywords{}

\maketitle

\section{Introduction}
Understanding the dynamics of quantum many-body systems is 
a central objective of modern physics which has been reignited by  
experimental advancements featuring, e.g., cold atoms or trapped ions 
\cite{Bloch2012,Blatt2012}, and has experienced an upsurge of interest also 
from the theoretical side \cite{Polkovnikov2011, Eisert2015, Dalessio2016, 
Gogolin2016, Borgonovi2016}. In this context, an intriguing and fundamental 
direction of research is to explain if and how thermodynamic behavior can emerge 
from the unitary time evolution of isolated quantum systems. One notable 
explanation for this occurrence of thermalization is 
the eigenstate thermalization hypothesis (ETH) \cite{Deutsch1991,Srednicki1994,Rigol2008}, 
which has been 
numerically verified in numerous instances \cite{Dalessio2016}. 

However, despite thermalization certainly being a rather common 
observation, there are 
also classes of systems which generically evade to reach thermal 
equilibrium even at 
indefinitely long times. In particular, it has been realized 
early on by Anderson that non-interacting particles in one or two spatial 
dimensions localize for an arbitrarily weak disorder potential 
\cite{Anderson1958,Abrahams1979} (for experimental confirmations see, e.g., 
\cite{Roati2008,Billy2008}).   
Moreover, it is now widely believed that for sufficiently strong 
disorder, localization is also possible in the presence of interactions  
\cite{Nandkishore2015,Abanin2019}, which is supported by 
experimental results as well \cite{Schreiber2015}. 

While the majority of studies on many-body localization (MBL) typically focus 
on one-dimensional and 
short-ranged models composed of, e.g., spin-$1/2$ degrees of freedom, there has 
been much effort recently to generalize the notion of MBL to a wider 
class of models \cite{Gopalakrishnan2019}. This includes, e.g., systems which are weakly coupled 
to a thermal bath \cite{Luschen2017}, models with long-range interactions \cite{Smith2016} or  
degrees of freedom with higher spin $S > 1/2$ \cite{Richter2019g,Richter2020}, as well as 
Hubbard models where the 
disorder only couples to either one of the charge or spin degrees of freedom 
\cite{Prelovsek2016,Kozarzewski2018}. 

A particularly interesting question is whether MBL can also occur 
in systems which are translational invariant, i.e.,  without any 
explicit disorder 
\cite{Carleo2012,DeRoeck2014,DeRoeck2014a,DeRoeck2014b,Grover2014,Schiulaz2014,
Altman2015,Papic2015,Schiulaz2015,Hickey2016,Yao2016,Smith2017,
Michailidis2018,Brenes2018,Sirker2019}.
A convenient model to investigate this question is given by the 
mass-imbalanced Hubbard chain \cite{Schiulaz2014,Grover2014,Gamayun2014,
Altman2015,Jin2015,Sirker2019}. In this model, 
two mutually interacting particle species
are defined on a one-dimensional lattice and exhibit different hopping 
amplitudes. Here, the imbalance is parametrized by the ratio $\eta$ between the 
two hopping strengths, ranging from ${\eta = 0}$, where the heavy particles 
are entirely static, to ${\eta = 1}$, where the hopping amplitudes are the same. 
On the one hand, in the balanced limit ${\eta = 1}$, numerical evidence for 
diffusive \cite{Prosen2012,Karrasch2017,Steinigeweg2017} (or superdiffusive 
\cite{Prosen2012,Ilievski2018}) charge transport has been found in the regime of 
high temperatures and strong interactions.
On the other hand, for ${\eta=0}$, the static particle species creates an 
effective disorder potential which induces localization of the lighter 
particles \cite{Paredes2005,Andraschko2014,Zhao2016,Enss2017,Jin2015}.
In view of these two opposing cases, it is intriguing to 
study the dynamics in the regime of  intermediate imbalances ${0 < \eta < 1}$.
While genuine localization (i.e.\ on indefinite time scales) is most likely 
absent for any ${\eta > 0}$ \cite{Yao2016,Sirker2019}, e.g., due to slow 
anomalous diffusion which ultimately leads to thermalization 
\cite{Yao2016}, this does not exclude the possibility of interesting dynamical 
properties such as a ``quasi-MBL phase'' at short to intermediate times
\cite{Yao2016}. 

In this paper, we scrutinize the impact of a finite mass imbalance ${0 < \eta < 
1}$ from a different perspective by studying the real-time dynamics of pure 
states featuring a sharp initial non-equilibrium density profile. 
 Relying on the concept of dynamical 
 quantum typicality, the resulting non-equilibrium dynamics
 can be related to equilibrium correlation functions.
 Summarizing our main results, we observe that 
 diffusive transport occurs for moderate 
 values of the mass imbalance, and that it manifests itself in a 
 Gaussian spreading of real-space density profiles and an exponential decay of 
 density modes in momentum space. Moreover, for stronger 
 imbalances, we find evidence that on the time and length scales numerically 
accessible, transport properties become anomalous, albeit we cannot rule out 
that normal diffusion eventually prevails at even longer times. 
Furthermore, our results are consistent with the absence of genuine 
localization for any ${\eta > 0}$.  
In particular, we find that for smaller and smaller values of ${\eta > 0}$, the 
resulting dynamics resembles the localized ${\eta = 0}$ case for longer 
and longer time scales. However, we conjecture that this ``lifetime'' of 
effective localization always remains finite for a finite $\eta$.     

 This paper is structured as follows. After introducing the model in 
 Sec.~\ref{sec:Model}, we give an introduction to the employed typicality approach, 
 and the initial states and observables in Sec.~\ref{sec:Method}. We then present 
 our results in Sec.~\ref{sec:Results} and conclude with a discussion in 
 Sec.~\ref{sec:Conclusion}.

\section{Model} \label{sec:Model}

\begin{figure}[tb]
	\centering
		\includegraphics[width=.45\textwidth]{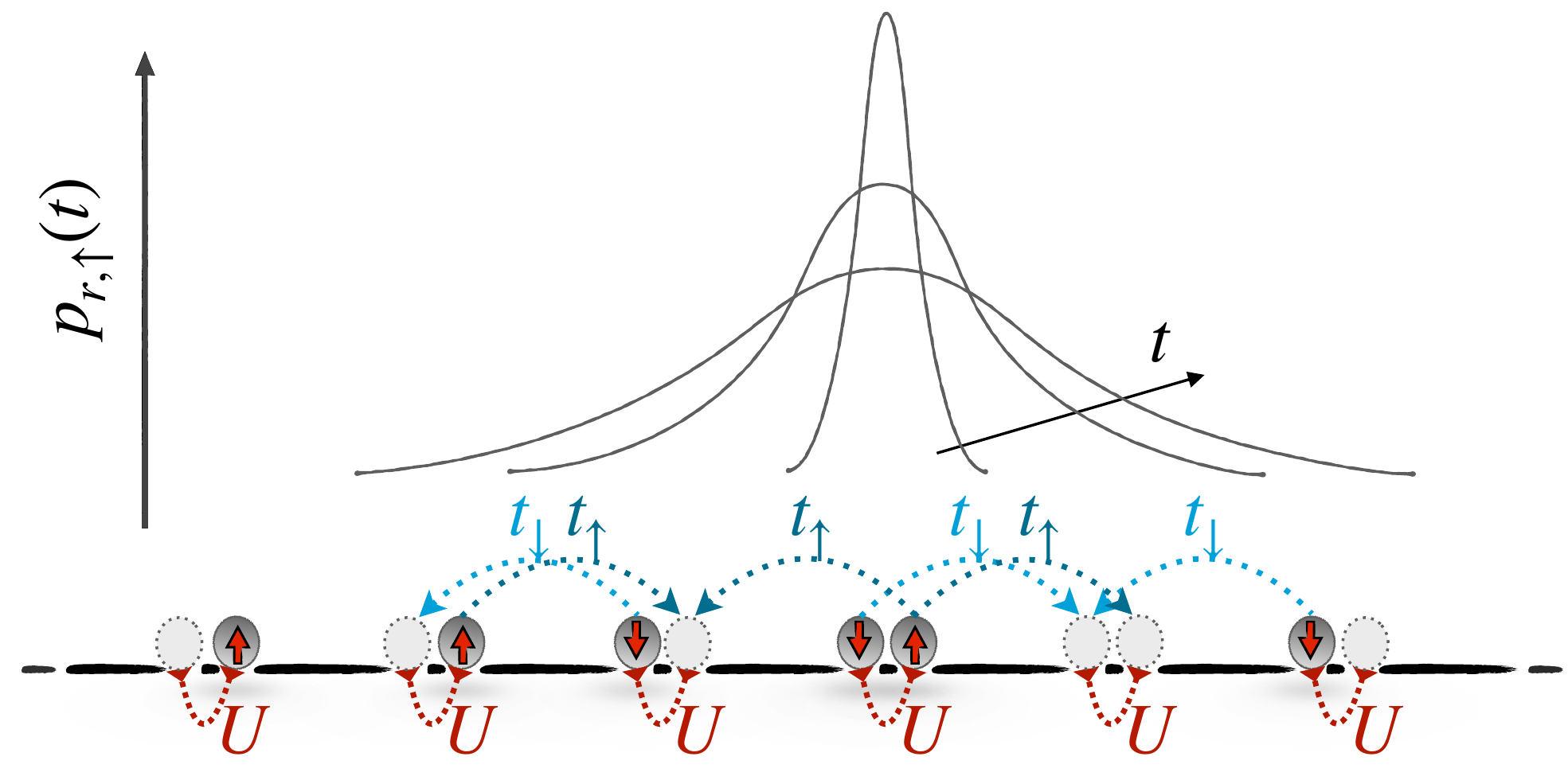}
	\caption{(Color online) Illustration of the imbalanced Fermi-Hubbard chain.
	Spin-$\uparrow$ and -$\downarrow$ particles with on-site interaction of strength $U$ and different hopping amplitudes $t_{\uparrow}$ and $t_{\downarrow}$. Diffusive broadening of 
	the initially peaked spin-$\uparrow$ density profile is sketched as a possible scenario depending on the imbalance ratio ${\eta=t_{\downarrow}/t_{\uparrow}}$.}
	\label{fig:model}
\end{figure}
We study the Hubbard chain describing interacting spin-$\uparrow$ and -$\downarrow$ 
fermions on a one-dimensional lattice.
The Hamiltonian for $L$ lattice sites   
with periodic boundary conditions (${L+1\equiv 1}$) reads
\begin{align}\label{eq:LatticeHamiltonian}
H=\sum\limits_{r=1}^{L}h_r
\end{align}
with local terms
\begin{align}\label{eq:local-hamilton-hubbard}
h_r=-\sum_{\sigma=\uparrow,\downarrow}&
t_{\sigma}\left(c_{r\phantom{l}\negmedspace,\sigma}^{\dag}c_{r+1,\sigma}^{\phantom\dag}
+\mathrm{h.c.}\right)\\\notag
+&\,U\left(n_{r,\uparrow}-\frac{1}{2}\right)\left(n_{r,\downarrow}-\frac{1}{2}\right)\ ,
\end{align}
where the creation (annihilation) operator $c_{r,\sigma}^{\dag}$ $(c_{r,\sigma})$ creates 
(annihilates) a fermion with spin $\sigma$ at site $r$,
and ${n_{r,\sigma}=c_{r,\sigma}^{\dag}c_{r,\sigma}^{\phantom\dag}}$ is the 
particle-number operator. (We omit any additional operator symbols for the sake of clean notation.) The first term on the 
right hand side of Eq.~\eqref{eq:local-hamilton-hubbard} describes the 
site-to-site hopping of each particle species with amplitude $t_{\sigma}$.
The second term is the on-site interaction between the particle species with 
strength $U$, see Fig.\ \ref{fig:model}. 
The imbalance between $t_{\uparrow}$ and $t_{\downarrow}\leq t _{\uparrow}$ 
is parametrized by the ratio
\begin{align}
\eta=\frac{t_{\downarrow}}{t_{\uparrow}}\ ,
\end{align}
ranging from ${\eta=0}$ for ${t_{\downarrow} = 0}$ to 
${\eta=1}$ in the case of ${t_\downarrow = 
t_\uparrow}$.

While the Hamiltonian $H$ in Eqs.\ \eqref{eq:LatticeHamiltonian} 
and \eqref{eq:local-hamilton-hubbard} is integrable in terms of the Bethe 
Ansatz for ${\eta = 1}$ (i.e.\ in the case of the standard Fermi-Hubbard chain, see, e.g., 
Ref.~\cite{Essler2005}), this 
integrability is broken for any finite imbalance ${0 < \eta < 1}$. 
Moreover, despite its integrability, there has been 
numerical evidence that, in the regime of high temperatures and strong 
interactions, charge transport in the one-dimensional Fermi-Hubbard model is 
diffusive \cite{Prosen2012,Karrasch2017,Steinigeweg2017} (or superdiffusive 
\cite{Prosen2012,Ilievski2018}). 
In order to have this well-controlled point of reference 
for our analysis of finite imbalances ${\eta \leq 1}$, we here fix the 
interaction strength to the large value ${U/t_{\uparrow} = 16}$.

In addition to ${\eta = 1}$, another important point in parameter 
space is the so-called Falicov-Kimball limit ${\eta=0}$ \cite{Falicov1969,Lyzwa1994}.
In this limit, the spin-$\downarrow$ 
particles become completely immobile (${t_{\downarrow}=0}$), implying 
that the local occupation numbers $n_{r,\downarrow}$ become strictly 
conserved quantities, i.e.,
\begin{align}
\left[H,n_{r,\downarrow}\right]=\left[n_{r,\downarrow},n_{r,\uparrow}\right]=0\ 
.
\end{align}
Using this symmetry, the Hamiltonian \eqref{eq:LatticeHamiltonian} can be 
decoupled into $2^L$ independent subspaces, effectively describing 
non-interacting spin-$\uparrow$ particles on a one-dimensional lattice with random 
(binary) on-site potentials ${\pm\,(U/2)\left(n_{r,\uparrow}-1/2\right)}$, which implies 
the onset of Anderson localization \cite{Anderson1958}.

It is worth mentioning that by means of a
Jordan-Wigner transformation, the fermionic model in 
Eqs.~\eqref{eq:LatticeHamiltonian} and 
\eqref{eq:local-hamilton-hubbard} can be mapped to a spin-$1/2$ 
model with ladder geometry \cite{Yao2016}. This spin model is described by the 
local terms 
\begin{align}\label{eq:local-hamilton-heisenberg}
h_r=-\sum_{k=1,2}&2J_k\left(s_{r\phantom{l}\negmedspace,k}^x s_{r+1,k}^x+s_{r\phantom{l}\negmedspace,k}^y s_{r+1,k}^y \right)\\\notag
+&\,J_{\bot }\,s_{r,1}^zs_{r,2}^z\ 
\end{align}
with ${J_{1}=t_{\uparrow}}$, ${J_2=t_{\downarrow}}$ and ${J_{\bot}=U}$.
Here, the different particle species $\uparrow$ and 
$\downarrow$ are represented as local magnetizations on 
the two separate legs ${k = 1,2}$
of the ladder. 
The hopping term and the interaction term in the Hubbard formulation correspond 
to the $XY$ interaction along the legs and the Ising interaction on the rungs 
of the ladder, respectively.
The particle number conservation ${N_{\sigma}=~\sum_r 
n_{r,\sigma}=~\mathrm{const.}}$ for both particle types 
translates into 
magnetization conservation ${M_{k}=~\sum_r s^z_{r,k}=~\mathrm{const.}}$ 
on each leg.

\section{Setup and numerical method} \label{sec:Method}

\subsection{Initial states and observables}

We investigate the real-time dynamics of local particle densities 
given by the expectation values 
\begin{align}\label{eq:density operator}
p_{r,\sigma}(t)=\Tr{n_{r,\sigma}\rho(t)}
\end{align}
with the density matrix
\begin{align}
\rho(t)=\mathrm{e}^{-\inice Ht}\ket{\psi(0)}\bra{\psi(0)}\mathrm{e}^{\inice Ht}
\end{align}
for pure initial states $\ket{\psi(0)}$, such that
\begin{align}\label{eq:scalar product}
	\Tr{n_{r,\sigma}\rho(t)}=\bra{\psi(t)}n_{r,\sigma}\ket{\psi(t)}\ .
\end{align}
In order to realize inhomogeneous particle densities, 
we prepare the initial states via the projection
\begin{align}\label{eq:initial states}
\ket{\psi(0)}\propto n_{L/2,\uparrow}\ket{\phi}\ .
\end{align}
The reference pure state $\ket{\phi}$ is constructed as 
a random superposition,
\begin{align}\label{eq:typical-state}
	\ket{\phi}=\sum\limits_{k=1}^{d}c_k\ket{\varphi_k} \ ,
\end{align}
where the $\ket{\varphi_k}$ denote the common 
eigenbasis of the local occupation number operators $n_{r,\sigma}$, 
and the sum runs over the full Hilbert space with finite dimension 
${d=4^L}$. 
(In spin language, this simply is the Ising basis.) 
Moreover, the complex coefficients $c_k$ are randomly drawn from a 
distribution which is invariant under all unitary transformations in the Hilbert 
space (Haar measure) \cite{Bartsch2009, Bartsch2011}, i.e., real and 
imaginary parts of these coefficients are normally distributed with zero mean.
As a consequence, the initial
density profile exhibits a sharp delta peak for the 
spin-$\uparrow$ particles in the middle of the chain on top of a homogeneous 
many-particle background \cite{Steinigeweg2017,Steinigeweg2017a},
\begin{align}\label{eq:initial profile}
p_{r,\sigma}(0)\ \begin{cases}
\quad =\phantom{1}1 & r=L/2 \ \mathrm{and}\ \sigma=\ \uparrow\\
\quad \approx1/2=\peq & \mathrm{else}
\end{cases} \ .
\end{align}
Rather than taking the full Hilbert space into account, one 
could also consider the half-filling sector (respectively the 
zero-magnetization sector). 

\subsection{Dynamical quantum typicality}

Given the specific construction of the pure state $\ket{\phi}$ 
in Eq.\ \eqref{eq:typical-state}, the concept of dynamical quantum typicality (DQT)
provides a direct connection between the non-equilibrium 
expectation value $p_{r,\uparrow}(t)$ and an 
equilibrium correlation function (see Ref.\ \cite{Steinigeweg2017} and also 
Appendix~\ref{app:typ-approx}), 
\begin{align}\label{eq:typicality-approximation}
p_{r,\uparrow}(t)-\peq=&\,2\braket{(n_{L/2,\uparrow}-\peq)(n_{r,\uparrow}(t)-\peq)}+\epsilon\ ,
\end{align}
where the thermodynamic average $\braket{\bullet}=\Tr{\bullet}/d$ is carried out at 
formally infinite temperature. As a 
consequence, the dynamics of the non-equilibrium expectation value 
$p_{r,\uparrow}(t)$ can be used to study transport properties within the 
framework of linear response theory.

Importantly, the variance of the statistical error ${\epsilon=\epsilon(\ket{\phi})}$ of Eq.\ 
\eqref{eq:typicality-approximation} is bounded from above by 
\begin{align}
\text{Var}(\epsilon)<\mathcal{O}\left(\frac{1}{d}\right)=\mathcal{O}\left(4^{-L}\right)\ ,
\end{align}
 i.e., the accuracy of the typicality 
approximation improves exponentially upon increasing the size of the 
system. In principle, this 
error can be further reduced by averaging over multiple 
realizations of the random state $\ket{\phi}$ \cite{Schnack2020,Schnack2020a}. However, for the 
system sizes studied here, the DQT approach is already very accurate, and this 
additional sampling becomes unnecessary \cite{Heitmann2020}.
More details on the concept of dynamical quantum typicality 
(and on error bounds) can be found in 
Refs.~\cite{Gemmer2003,Iitaka2003,Iitaka2004,Goldstein2006,Popescu2006, 
Reimann2007,White2009,Sugiura2012,Sugiura2013,Elsayed2013,Steinigeweg2014, 
Steinigeweg2014a,Monnai2014,Reimann2016,Reimann2018,Heitmann2020}.

\subsection{Time evolution via pure-state propagation}

For the time evolution of the pure state 
\begin{align}
	\ket{\psi(t)}=\mathrm{e}^{-\inice Ht}\ket{\psi(0)}
\end{align} 
we can bypass the exact diagonalization of the Hamiltonian and rather 
solve the time-dependent Schrödinger equation directly via iterative 
forward propagation in small time steps $\delta t$.
Aside from the many numerical methods such as Trotter decompositions 
\cite{DeVries1993,DeRaedt2006}, Chebyshev polynomials 
\cite{TalEzer1984,Dobrovitski2003,Weisse2006} or Krylov-space methods 
\cite{Nauts1983}, the action of the time-evolution operator in each 
step can be calculated by a fourth-order Runge-Kutta scheme 
\cite{Elsayed2013,Steinigeweg2014},
\begin{align}\label{eq:RK4}
	\ket{\psi(t+\delta t)}&=\mathrm{e}^{-\inice H\delta t}\ket{\psi(t)}\\\notag
						 &\approx\sum\limits_{k=0}^4\frac{(-\inice H\delta t)^k}{k!}\ket{\psi(t)} \ .
\end{align}
Crucially, the matrix-vector multiplications in Eq.~\eqref{eq:RK4} 
can be implemented very memory efficiently due to the sparse matrix representation 
of the given Hamiltonian.  
While the action of $H$ on $\ket{\psi}$ can also be 
calculated on-the-fly, we save the sparse Hamiltonian matrix for the sake of run time.
Moreover, symmetries of the system can be exploited 
in order to split the problem into smaller sub-problems and to further reduce the 
computational effort
\cite{Heitmann2019}. 
In this paper, we exploit the particle number (magnetization) 
conservation for both particle species (legs) separately. As a 
consequence, the maximum memory consumption for the largest symmetry sector in a 
system of length ${L=15}$, 
with full Hilbert-space dimension ${d\sim 
10^9}$, amounts to about $20$ GB (using double-precision complex numbers). 
While ${L=14}$ or ${L=15}$ are already comparatively large (especially in view of the extensive parameter screening and the long simulation times considered here), let us note that even larger system sizes can be treated by the usage of large-scale supercomputing (see, e.g., Refs.~\cite{Jin2015,Steinigeweg2017}).
The time step used in all calculations, if not stated 
otherwise, is ${\delta \tth=0.005}$.

\subsection{Diffusion on a lattice}

\subsubsection{Real space}

The dynamics of the densities $p_{r,\uparrow}(t)$ is 
diffusive, if it fulfills the lattice diffusion equation \cite{Bertini2020} 
\begin{align}\label{eq:diffusion-equation}
	\frac{d}{dt}p_{r,\uparrow}(t)=D[p_{r-1,\uparrow}(t)-2p_{r,\uparrow}(t)+p_{r+1,\uparrow}(t)]
\end{align}
with the diffusion constant $D$. 
For the $\delta$-peak initial conditions \eqref{eq:initial 
profile}, the solution of Eq.\ \eqref{eq:diffusion-equation} can be well 
approximated by the Gaussian function
\begin{align}\label{eq:gaussianfit}
	p_{r,\uparrow}(t)-\peq=\frac{1}{2\sqrt{2\pi}\Sigma(t)}\exp\left[-\frac{(r-L/2)^{2}}{2\Sigma^2(t)}\right]\ ,
\end{align}
where the spatial variance scales as ${\Sigma^2(t)=2Dt}$ and is given by 
\begin{align}\label{eq:spatial-variance}
\Sigma^2(t)=\sum\limits_{r=1}^Lr^2\delta p_{r,\uparrow}(t)-\left[\sum\limits_{r=1}^L r\delta p_{r,\uparrow}(t)\right]^2\ ,
\end{align}
with ${\delta p_{r,\uparrow}(t)\propto p_{r,\uparrow}(t)-\peq}$ fulfilling ${\sum_r \delta p_{r,\uparrow}(t)=1}$ for all times $t$.
More generally, a 
scaling of the variance according to ${\Sigma(t)\propto t^{\alpha}}$ is called 
ballistic for ${\alpha = 1}$, superdiffusive for ${1/2 < \alpha < 1}$, 
diffusive for ${\alpha = 1/2}$, subdiffusive for ${0< \alpha < 1/2}$, and localized 
for ${\alpha = 0}$. 
Moreover, away from the case ${\alpha=1/2}$, the density profiles $p_{r,\uparrow}(t)$ 
are not expected to take on a Gaussian shape.

\subsubsection{Connection to current-current correlation 
functions}

Due to the typicality relation 
\eqref{eq:typicality-approximation}, the spatial variance in Eq.~\eqref{eq:spatial-variance}
can be related to the dynamics of current-current 
correlation functions via \cite{Steinigeweg2009b}  
\begin{align}\label{eq:variance relation}
	\frac{d}{dt}\Sigma^2(t)=2D(t) \ ,
\end{align}
where the time-dependent diffusion coefficient $D(t)$ is given by
\begin{align}\label{eq:diff-const}
	D(t)=\frac{4}{L}\int\limits_0^t\langle 
j_{\uparrow}(t')j_{\uparrow}\rangle\,dt'\ ,
\end{align} 
and $j_{\uparrow}$ denotes the total current operator of 
the spin-$\uparrow$ particles,
\begin{align}
	j_{\uparrow}=-t_{\uparrow}\sum\limits_r\left(\ \inice c_{r,\uparrow}^{\dag}c_{r+1,\uparrow}^{\phantom\dag}+\hc\right)\ .
\end{align}
(Note that the relation \eqref{eq:variance relation} requires $\delta p_{r,\uparrow}(t)$
to vanish at the boundaries of the chain \cite{Steinigeweg2009b}.)
We therefore can compare the spatial variance of density 
profiles calculated according to Eq.~\eqref{eq:spatial-variance} to the one already 
obtained from current-current correlation functions
\cite{Steinigeweg2009b,Yan2015,Karrasch2017},
\begin{align}\label{eq:sigmaLR}
	\Sigma^2(t)=2\int\limits_0^{t}D(t')dt'\ .
\end{align} 
A detailed analysis of transport in the mass-imbalanced Hubbard chain extracted 
from current-current correlation functions can be found in \cite{Jin2015}.

\subsubsection{Momentum space}
 In addition to the real-space perspective, it is also instructive to look at 
 momentum-space observables as given by the lattice Fourier transform of the density profiles, 
\begin{align}
	p_{q,\uparrow}(t)=\frac{1}{\sqrt{L}}\sum\limits_{r=1}^L\mathrm{e}^{\inice q r}p_{r,\uparrow}(t)
\end{align}
with the momentum ${q=2\pi k/L}$ and wave numbers ${k=0,1,\dots,L-1}$.
In particular, the Fourier transformation of the diffusion equation 
\eqref{eq:diffusion-equation} yields the corresponding diffusion equation 
for the $p_{q,\uparrow}(t)$,
\begin{align}\label{Eq::DiffMom}
	\frac{d}{dt}p_{q,\uparrow}(t)=-\widetilde{q}^2D\,p_{q,\uparrow}(t)\ ,
\end{align}
with ${\widetilde{q}^2=2(1-\cos q)}$.
From Eq.\ \eqref{Eq::DiffMom}, it becomes clear that diffusion manifests 
itself in momentum space by exponentially decaying modes 
\begin{align}\label{eq:exp-decay-fourier}
	p_{q,\uparrow}(t)\propto\emath{-\widetilde{q}^2Dt}\ .
\end{align} 

\section{Results} \label{sec:Results}

We now turn to our numerical results. To begin with, the two limiting cases ${\eta = 1}$ 
and ${\eta = 0}$ are presented in Sec.~\ref{sec:Results-Diff-vs-Loc}. 
Intermediate imbalances ${0 < \eta < 1}$ are discussed
in Secs.\ \ref{sec:Results-robust-diff} and \ref{sec:Results-localization}.

\subsection{Limiting cases}\label{sec:Results-Diff-vs-Loc}

\begin{figure}[tb]
	\centering
		\includegraphics[width=.45\textwidth]{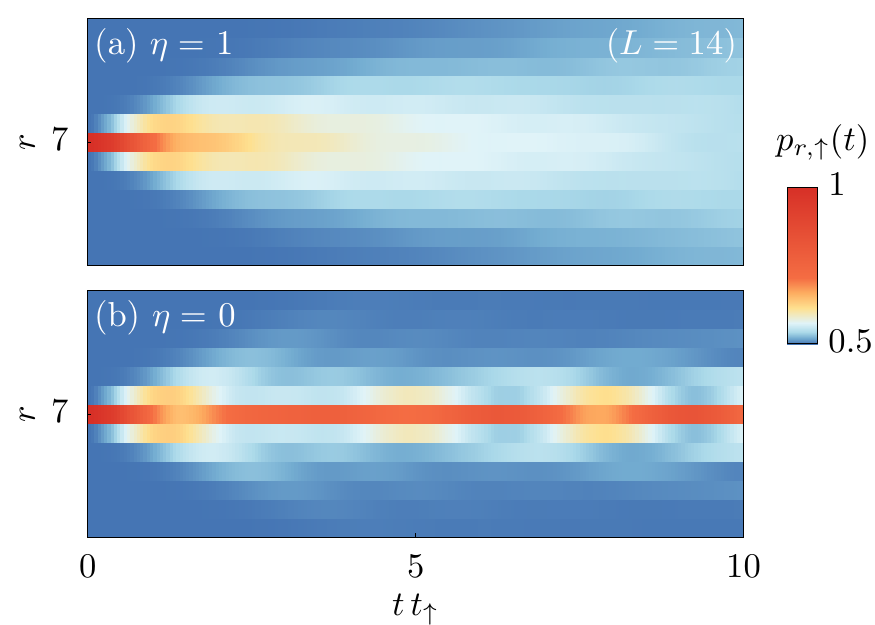}
	\caption{(Color online) Real-time broadening of the non-equilibrium density profile 
	for limiting imbalance ratios (a) $\eta=1$ and (b) $\eta=0$. System size $L=14$ 
	and interaction strength $U/t_{\uparrow}=16$. The initial density peak in the 
	center of the chain spreads rather quickly over the system for $\eta=1$, 
	whereas it appears to be frozen for $\eta=0$.}
	\label{Fig1}
\end{figure}
\begin{figure}[b]
	\centering
		\includegraphics[width=.45\textwidth]{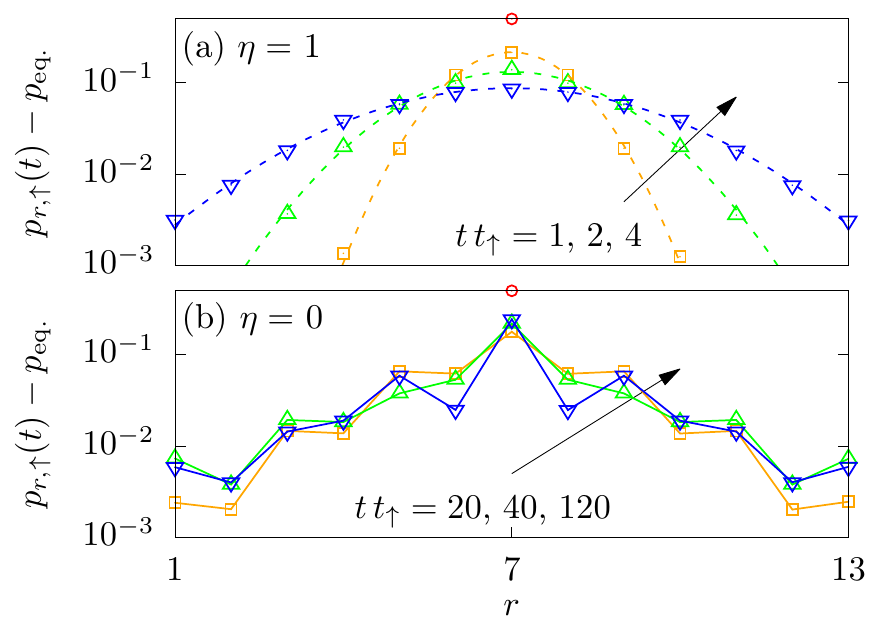}
	\caption{(Color online) Density profiles $p_{r,\uparrow}(t)$ at fixed times for 
	(a) $\eta=1$ and (b) $\eta=0$. In the case $\eta=1$, the profiles can be very 
	well described via Gaussians (parabola in the semi-logarithmic plot) indicating 
	clean diffusion for the time scales depicted. Note that the Gaussians (dashed lines) are no fit, but calculated from Eqs.~\eqref{eq:gaussianfit} and \eqref{eq:spatial-variance}. In the case $\eta=0$, an overall 
	triangular shape survives even for long times, with some local fluctuations.}
	\label{Fig2}
\end{figure}

In order to mark out the two completely different behaviors of the density dynamics 
in the limiting cases of the model, we first discuss the limit of equal particle 
masses ($\eta=1$) and contrast it with the limit of infinite mass-imbalance (${\eta=0}$). 
Recall that the interaction strength is set to 
${U/t_{\uparrow}=16}$ in the following.

First, \figref{Fig1} shows the real-time broadening of the initially peaked 
density profiles $p_{r,\uparrow}(t)$ for both limits in a time-space density plot. 
While the particle density for $\eta=1$ [\figref{Fig1}~(a)] is found to spread over 
all sites of the chain, $p_{r,\uparrow}(t)$ for $\eta=0$ [\figref{Fig1}~(b)] 
appears to be essentially frozen at the central lattice sites, as it is expected in 
the Anderson insulating limit.

For a more detailed analysis, the spatial dependence of the profiles ${p_{r,\uparrow}(t)-\peq}$ 
is shown in \figref{Fig2} for fixed times $t$ in a semi-logarithmic plot. 
Remarkably, the profiles for ${\eta=1}$ in \figref{Fig2}~(a) can be very well described by 
Gaussians [see Eq.~\eqref{eq:gaussianfit}] over three orders of magnitude. 
These Gaussian profiles indicate that charge transport in the integrable Fermi-Hubbard chain 
is diffusive \cite{Prosen2012,Karrasch2017,Steinigeweg2017}, at least in this parameter regime 
(strong interactions and high temperatures) and for the time scales depicted, see also 
Refs.~\cite{Prosen2012,Ilievski2018} for the possibility of 
superdiffusive transport.
Note that the Gaussians in \figref{Fig2}~(a) are no fit, since the width $\Sigma(t)$ has 
been calculated exactly according to Eq.~\eqref{eq:spatial-variance}, i.e., there is no 
free parameter involved. 
In contrast, the profiles for ${\eta=0}$ in \figref{Fig2}~(b) 
are clearly non-Gaussian and remain, 
even for the long times shown, in an overall triangular shape with variations on short 
length scales.

\subsection{Small imbalances}\label{sec:Results-robust-diff}
\subsubsection{Real space}
\begin{figure}[tb]
	\centering
		\includegraphics[width=.45\textwidth]{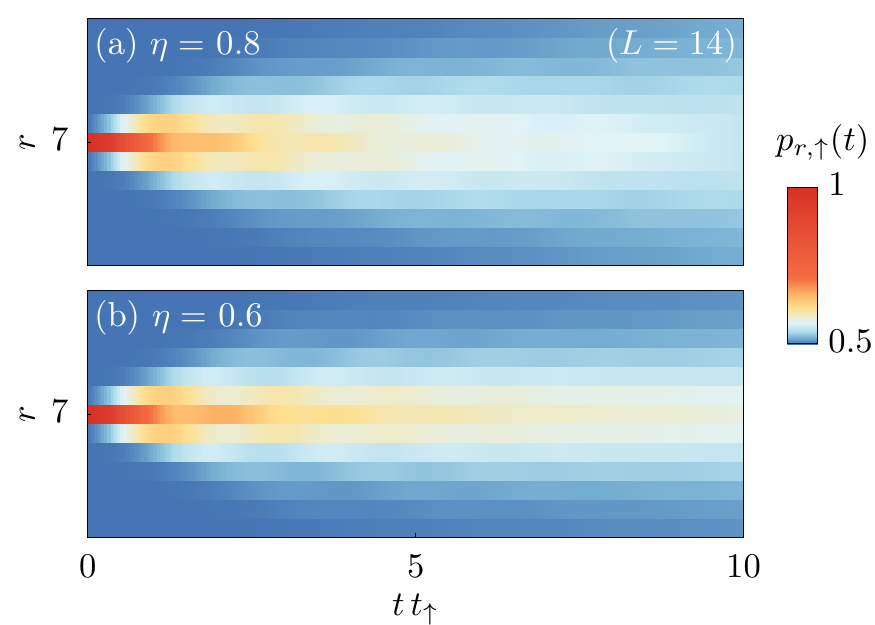}
	\caption{(Color online) Real-time broadening of the initially peaked density profile 
	for weak imbalances (a) $\eta=0.8$ and (b) $\eta=0.6$.}
	\label{fig:2dbroadening-imbalanced}
\end{figure}

\begin{figure}[tb]
	\centering
		\includegraphics[width=.45\textwidth]{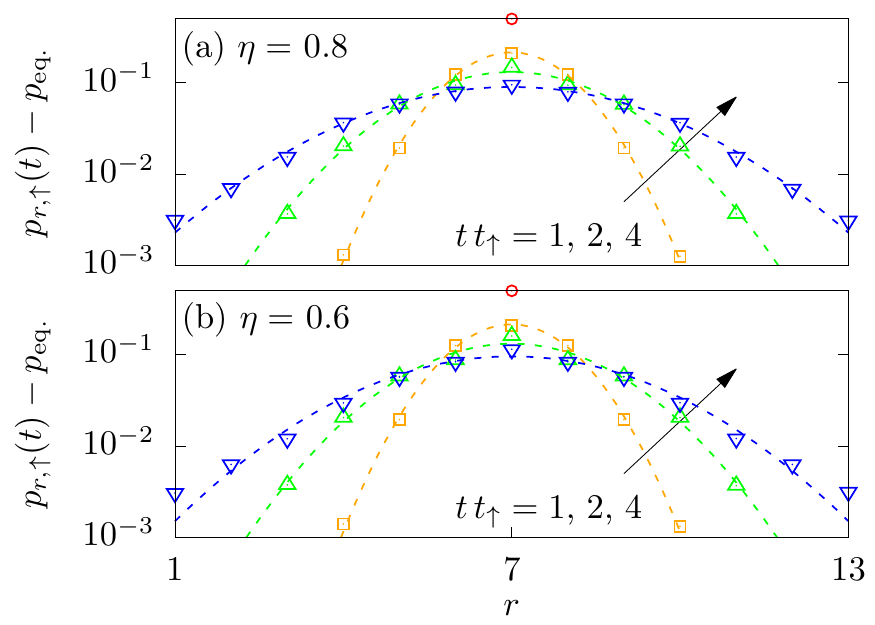}
	\caption{(Color online) Density profiles $p_{r,\uparrow}(t)$ at fixed times for the 
	same system parameters and imbalance ratios as shown in 
	\figref{fig:2dbroadening-imbalanced}. The dashed lines are Gaussian
	functions calculated from Eqs.~\eqref{eq:gaussianfit} and \eqref{eq:spatial-variance}. At moderate imbalance (a) ${\eta=0.8}$, the 
	profiles can be very well described by Gaussians. At slightly smaller (b) ${\eta=0.6}$, 
	the density profiles are still in good agreement with Gaussians, although small 
	deviations become apparent at ${\tth=4}$.}
	\label{fig:profiles-imbalanced}
\end{figure}

Next, let us study a finite imbalance between the particle masses. In analogy to \figref{Fig1}, 
time-space density plots are shown in \figref{fig:2dbroadening-imbalanced} for ${\eta=0.8}$ and 
${\eta=0.6}$. For these ratios the broadening of the initial density peak apparently happens on 
a time scale comparable to the one observed for ${\eta=1}$ in \figref{Fig1}, with a barely 
noticeable slowdown with the increasing imbalance. 
Similar observations can be made for the density profiles at fixed times, as  shown in 
\figref{fig:profiles-imbalanced}. At weak imbalance ${\eta=0.8}$ 
[\figref{fig:profiles-imbalanced}~(a)], the profiles are still in very good agreement 
with Gaussians [see Eq.~\eqref{eq:gaussianfit}] which suggests
that diffusion occurs also 
for ${\eta\neq1}$. Even for stronger imbalance ${\eta=0.6}$ [\figref{fig:profiles-imbalanced}~(b)], 
the profiles appear to be of Gaussian shape, although small deviations start to appear at 
${\tth=4}$, which might be seen as the onset of a drift from normal to anomalous diffusion, 
see also the discussion below.

\subsubsection{Spatial width}
\begin{figure}[b]
	\centering
		\includegraphics[width=.45\textwidth]{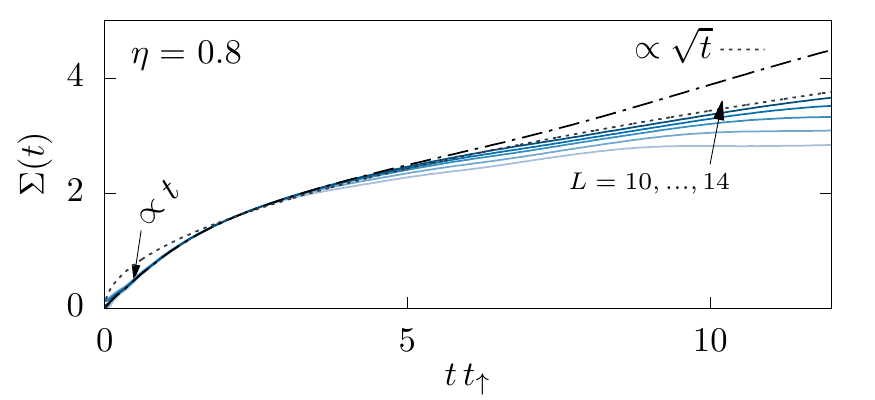}
	\caption{(Color online) Spatial width $\Sigma(t)$ as obtained by 
	Eq.~\eqref{eq:spatial-variance} for imbalance ratio ${\eta=0.8}$ and different 
	system sizes ${L=10,\dots,14}$ (arrow). The dotted line with the scaling 
	$\Sigma(t)\propto \sqrt{t}$ is a fit to the $L=14$ curve. 
	The width $\Sigma(t)$ (dashed-dotted line) as 
	calculated by Eq.~\eqref{eq:sigmaLR} is shown for comparison (${L=14}$).} 
	\label{Fig5}
\end{figure}
In order to analyze the broadening of the density profiles 
further, \figref{Fig5} shows the 
time-dependence of the spatial width 
$\Sigma(t)$ obtained by Eq.~\eqref{eq:spatial-variance} for 
moderate imbalance  
${\eta=0.8}$ and different system sizes ${L = 10,\dots,14}$.
Necessarily, there is an initial linear increase ${\Sigma(t)\propto t}$ for ${\tth\lesssim1}$, 
indicating ballistic transport, as it is expected for short times
below the mean-free time.
Subsequently, 
$\Sigma(t)$ shows a scaling 
${\propto\sqrt{t}}$, consistent with diffusion.
However, for later times, we find that $\Sigma(t)$ 
approaches a saturation value which increases 
with increasing $L$. 
This behavior of $\Sigma(t)$ can be easily understood since 
the width of a density profile on a finite lattice with $L$ sites is obviously bounded from 
above. Namely, assuming equilibration, i.e., a perfectly homogeneous distribution of 
the $p_{r,\uparrow}$ for $t\rightarrow\infty$ with $\delta p_{r,\uparrow}=1/L$ at each 
site, we obtain the saturation value
\begin{align}\label{eq:saturation}
	\Sigma^2(t\rightarrow\infty)&=\sum\limits_{r=1}^L\frac{r^2}{L}-\left[\sum\limits_{r=1}^L\frac{r}{L}\right]^2\\\notag
	&=\frac{1}{12}\left(L^2-1\right)\ .
\end{align}
This $L$-dependent saturation value is reached quickly for the weakly imbalanced case 
${\eta=0.8}$ in \figref{Fig5}, e.g., ${\Sigma \approx 4}$ for ${L=14}$.

Moreover, for the biggest size ${L=14}$, \figref{Fig5} also shows 
$\Sigma(t)$ calculated
from current-current correlation functions via Eq.~\eqref{eq:sigmaLR}.
Overall, the behavior of this $\Sigma(t)$ is 
in good agreement with the one 
described above. Note that the small deviations between the 
two widths setting in at ${\tth\sim6}$ 
presumably arise when the tails of the density distribution reach the boundaries of 
the system (cf. \figref{fig:profiles-imbalanced}).
Additionally, we note that the finite-size saturation value 
\eqref{eq:saturation} does not apply 
to Eq.~\eqref{eq:sigmaLR}, which, by definition, 
is not bounded. Rather, for times ${\tth\gtrsim6}$, we find an accelerated increase 
of $\Sigma(t)$. 
This is caused by the fact that the 
current-current correlation function ${\langle j_\uparrow(t) j_\uparrow 
\rangle}$ does not completely decay to zero in a system of finite size, 
see also Refs.\ \cite{Jin2015,Bertini2020}.

\subsubsection{Momentum space}
\begin{figure}[tb]
	\centering
		\includegraphics[width=.45\textwidth]{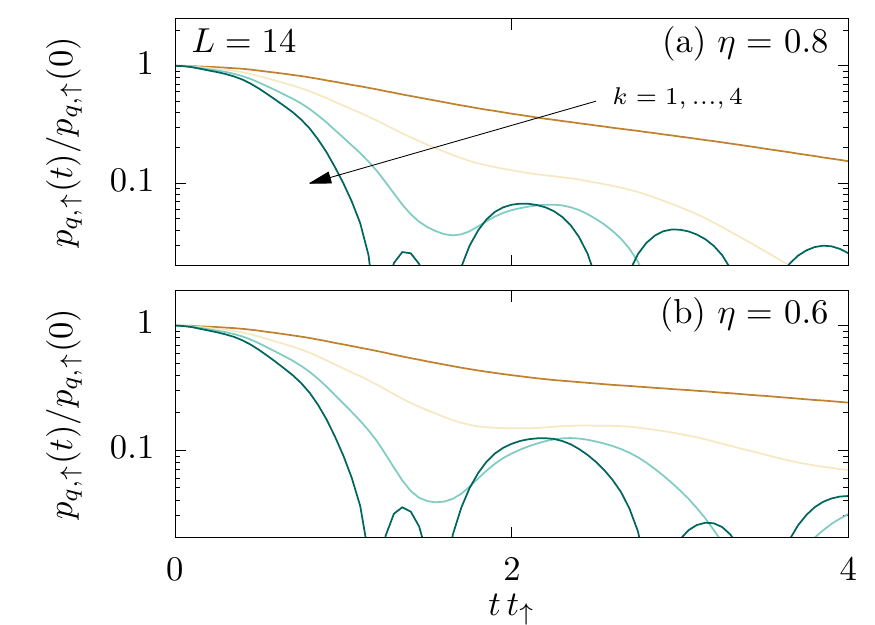}
	\caption{(Color online) Discrete Fourier transform $p_{q,\uparrow}(t)$ of the density 
	profile with momentum ${q=2\pi k/L}$ and wave numbers ${k=1,2,3,4}$ (arrow) for weak 
	imbalances (a) ${\eta=0.8}$ and (b) ${\eta=0.6}$.}
	\label{Fig8b}
\end{figure}
Complementary to the real-space data for ${\eta = 0.8,0.6}$
shown in 
\figref{fig:profiles-imbalanced}, the corresponding Fourier modes
$p_{q,\uparrow}(t)$ with momentum ${q=2\pi k/L}$ are shown in \figref{Fig8b} for the 
four longest wavelengths available, i.e., ${k=1,\dots,4}$. 
While $p_{q,\uparrow}(t)$ decays rather quickly for ${k \geq 2}$ 
(with the decay rate increasing with $k$), we find that at least for ${k = 1}$, 
$p_{q,\uparrow}(t)$ is to good quality described by an exponential decay [see 
Eq.~\eqref{eq:exp-decay-fourier}], consistent with the onset of diffusion on 
the corresponding length scales.

\subsection{Strong mass imbalance}\label{sec:Results-localization}
Now, let us study how the equilibration dynamics alter for stronger imbalances 
and also discuss the possibility of localization for ${\eta>0}$.

\subsubsection{Real-space dynamics}

\begin{figure}[b]
	\centering
		\includegraphics[width=.45\textwidth]{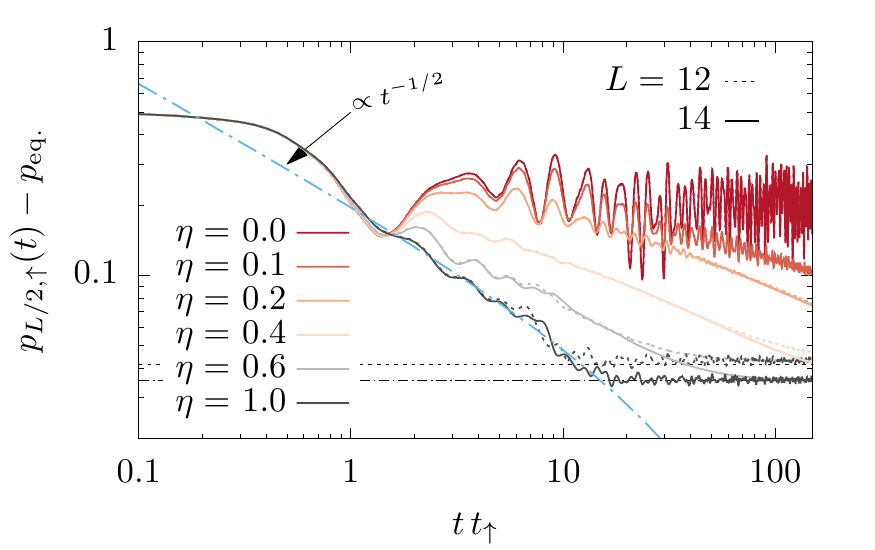}
	\caption{(Color online) Decay of the central peak $p_{L/2,\uparrow}(t)$ 
	at different imbalances ranging from ${\eta=0}$ to ${\eta=1}$ (from top to bottom) 
	for system sizes ${L=12}$ (dotted) and ${L=14}$ (solid). Dashed lines indicate the 
	expected $L$-dependent long-time value $(1-\peq)/L$.}
	\label{Fig-center}
\end{figure}

\begin{figure}[t]
	\centering
		\includegraphics[width=.45\textwidth]{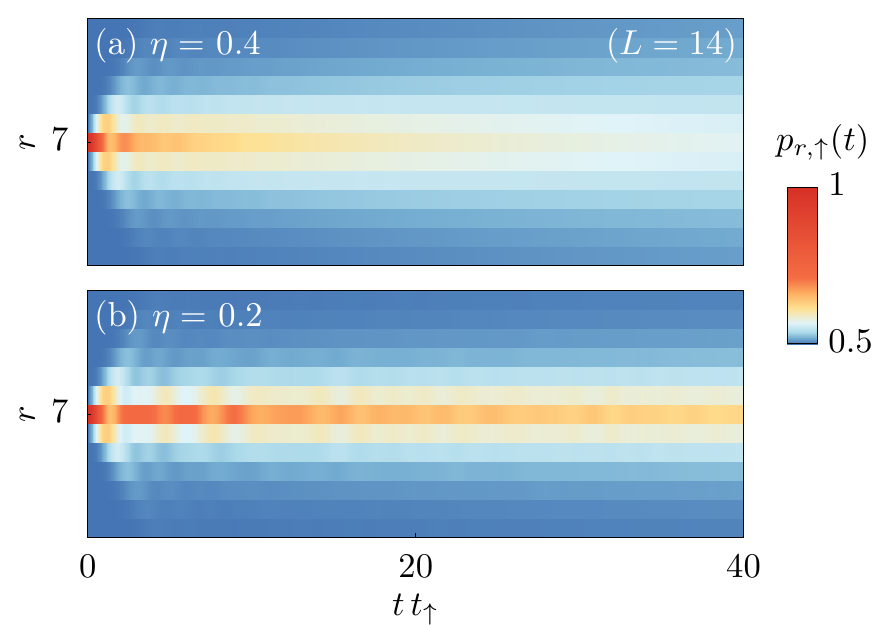}
	\caption{(Color online) Time-space density plot as in \figref{fig:2dbroadening-imbalanced} but for 
	(a) ${\eta=0.4}$ and (b) ${\eta=0.2}$.}
	\label{fig:2dbroadening-imbalanced-strong}
\end{figure}

\begin{figure}[b]
	\centering
		\includegraphics[width=.45\textwidth]{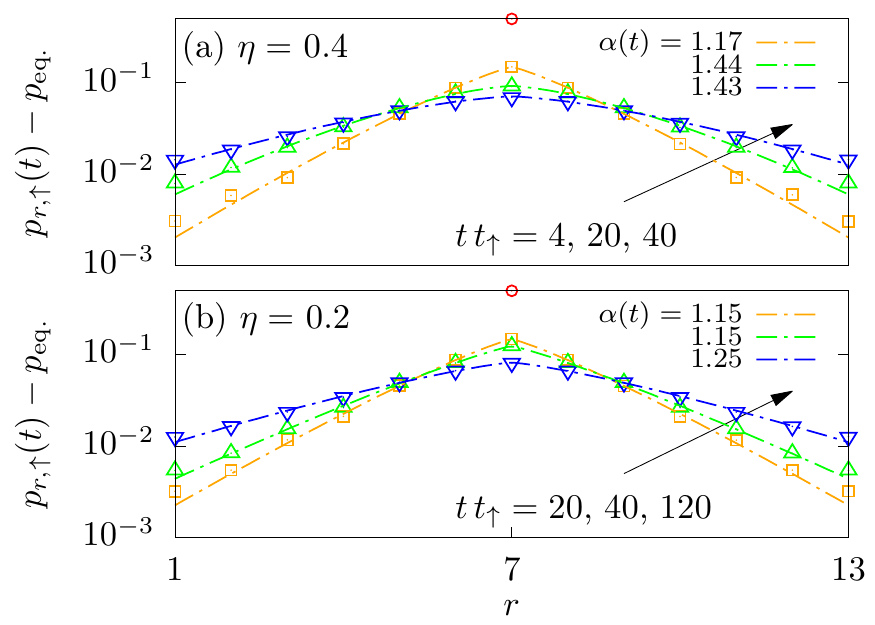}
	\caption{(Color online) Density profiles $p_{r,\uparrow}(t)$ as in \figref{fig:profiles-imbalanced} but for smaller 
	(a) ${\eta=0.4}$ and (b) ${\eta=0.2}$. The profiles broaden much slower and take on a 
	triangular shape in the semi-logarithmic plot used. The fit parameter $\alpha(t)$ is 
	the exponent used in Eq.~\eqref{eq:triangularfit}.}
	\label{fig:profiles-imbalanced-strong}
\end{figure}

Before discussing the full density profile in detail, let us for simplicity focus on the 
decay of the central peak $p_{L/2,\uparrow}(t)$, as shown in \figref{Fig-center} for 
imbalance ratios ${\eta=0,\dots,1}$ and two system sizes ${L=12}$ and ${L=14}$. 
While ${p_{L/2,\uparrow}(t) \propto t^{-1/2}}$ to good quality for $\eta =1$, 
consistent with diffusive transport, this decay is 
slowed down with decreasing $\eta$. 
At small but finite $\eta=0.1$, we find that $p_{L/2,\uparrow}(t)$ 
approximately coincides with the 
$\eta=0$ curve up to times ${\tth \approx 40}$, until it 
eventually starts to decay towards the equilibrium value
${p_{L/2,\uparrow}(t\rightarrow\infty)-\peq=(1-\peq)/L}$. 
Note that the two curves for ${L=12,14}$ agree very well with 
each other before the equilibration value is reached. On 
these time scales, the behavior of the density 
dynamics thus appears to be independent of the system size. This also illustrates the accuracy of the DQT approach, since there is no sign of sample-dependence in the time-dependent fluctuations 
of the strongly imbalanced curves.
For additional data with smaller $\eta$ and longer time scales, 
see Appendix~\ref{app:longtimes}. Moreover, a more detailed finite-size 
analysis can be found in Appendix~\ref{app:L-independence}.

Next, let us come back to a discussion of the full density profile. To this end, 
\figref{fig:2dbroadening-imbalanced-strong} shows 
time-space density plots for the two ${\eta=0.4}$ and ${\eta = 0.2}$. We find that the broadening 
of the density profiles visibly slows down with decreasing $\eta$, until 
no substantial spreading of the density can be observed for ${\eta=0.2}$ up to 
the maximum time ${\tth=40}$ shown here, consistent with \figref{Fig-center} discussed before. 

The corresponding cuts of the density profiles at fixed times are shown in Figs.\ 
\ref{fig:profiles-imbalanced-strong}~(a) and (b). Note that, owing to 
the slow broadening of the profiles, we show cuts at later times compared to Fig.\ 
\ref{fig:profiles-imbalanced}. One clearly sees that the 
profiles are not Gaussian anymore, but rather exhibit a pronounced 
triangular shape in the semi-logarithmic plot used. In particular, they can be well described 
by the function
\begin{align}\label{eq:triangularfit}
	p_{r,\uparrow}(t)-\peq=\beta(t)\exp\left[-\frac{\lvert r-L/2\rvert^{\alpha(t)}}{2\Sigma^2_{\mathrm{f}}(t)}\right]
\end{align}
with the time-dependent fit parameters $\alpha(t)$, $\Sigma_{\mathrm{f}}(t)$, and $\beta(t)$. In particular, 
the exponent ${\alpha(t) \in [1,2]}$ is introduced to capture the triangular shape. This shape 
indicates a crossover to anomalous diffusion for 
small ratios ${\eta \lesssim 0.4}$ \cite{Znidaric2016}. This is another central result of this 
paper.
\subsubsection{Spatial width}
\begin{figure}[t]
	\centering
		\includegraphics[width=.45\textwidth]{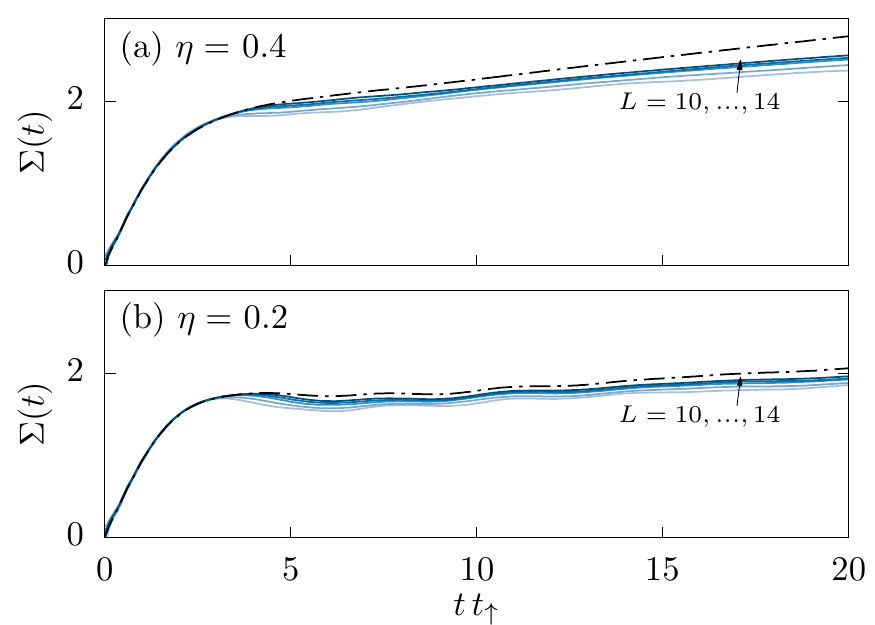}
	\caption{(Color online) Spatial width $\Sigma(t)$ for different system sizes ${L=10,\dots,14}$ 
	(arrows) as obtained by Eq.~\eqref{eq:spatial-variance} at (a) ${\eta=0.4}$ and (b) ${\eta=0.2}$. 
	The width $\Sigma(t)$ (dashed-dotted lines) as calculated according to Eq.~\eqref{eq:sigmaLR} is 
	shown for comparison (${L=14}$).} 
	\label{fig:variance-strong}
\end{figure}

Additionally, \figref{fig:variance-strong} shows the width $\Sigma(t)$ of the 
density profiles for ${\eta=0.4}$ and ${\eta=0.2}$, as calculated by Eqs.~\eqref{eq:spatial-variance} 
and \eqref{eq:sigmaLR}.
Compared to the weakly imbalanced case shown in \figref{Fig5}, $\Sigma(t)$ 
now grows much slower, 
and  Eqs.~\eqref{eq:spatial-variance} and \eqref{eq:sigmaLR} are in better agreement, since the 
distribution is still well concentrated in the center of the chain.
For ${\eta=0.2}$, $\Sigma(t)$ appears to remain at a constant plateau up to the 
maximum time ${\tth=20}$ shown.

\begin{figure}[t]
	\centering
		\includegraphics[width=.45\textwidth]{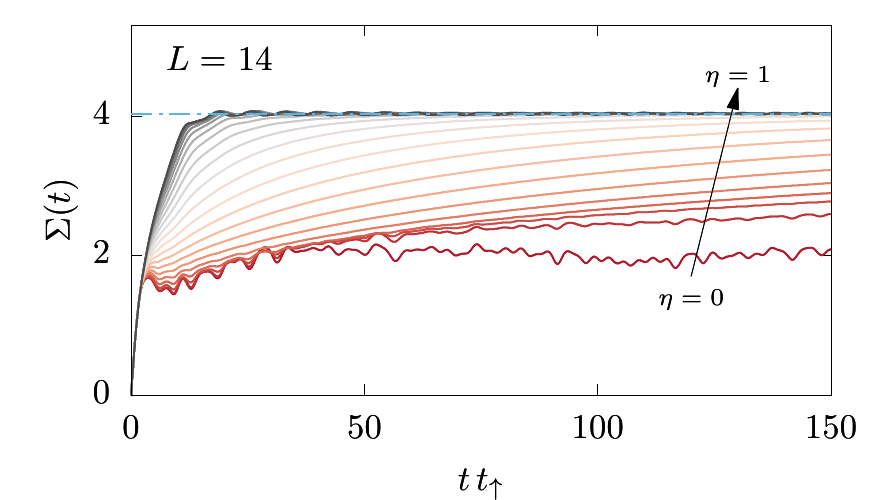}
	\caption{(Color online) Spatial width $\Sigma(t)$ for fixed system size $L=14$ and varying 
	imbalances ranging from ${\eta=0}$ to ${\eta=1}$ (arrow) in steps of 0.05. In the balanced case 
	${\eta=1}$, the width reaches its natural saturation value (dashed line) of ${\Sigma\approx 4}$ 
	[cf. Eq.~\eqref{eq:saturation}] rather quickly, while the other curves grow slower as $\eta$ 
	goes to zero. The curve for ${\eta=0}$ remains at around ${\Sigma\approx 2}$.}
	\label{Fig6}
\end{figure}

To analyze the $\eta$-dependence of the width in more detail, \figref{Fig6} shows 
$\Sigma(t)$ in Eq.~\eqref{eq:spatial-variance} on a longer time scale ${\tth\leq150}$ 
for various values of $\eta$ and a fixed system size ${L=14}$. While the growth 
of $\Sigma(t)$ towards the saturation value becomes slower and slower 
with decreasing $\eta$, we find that even for the smallest value of ${\eta = 
0.05}$ shown here, $\Sigma(t)$ clearly increases at long times. In contrast, 
the width in the ${\eta=0}$ case fluctuates around a constant and 
lower value, which might be interpreted as the Anderson localization length. 

\subsubsection{Momentum-space dynamics}

\begin{figure}[b]
	\centering
		\includegraphics[width=.45\textwidth]{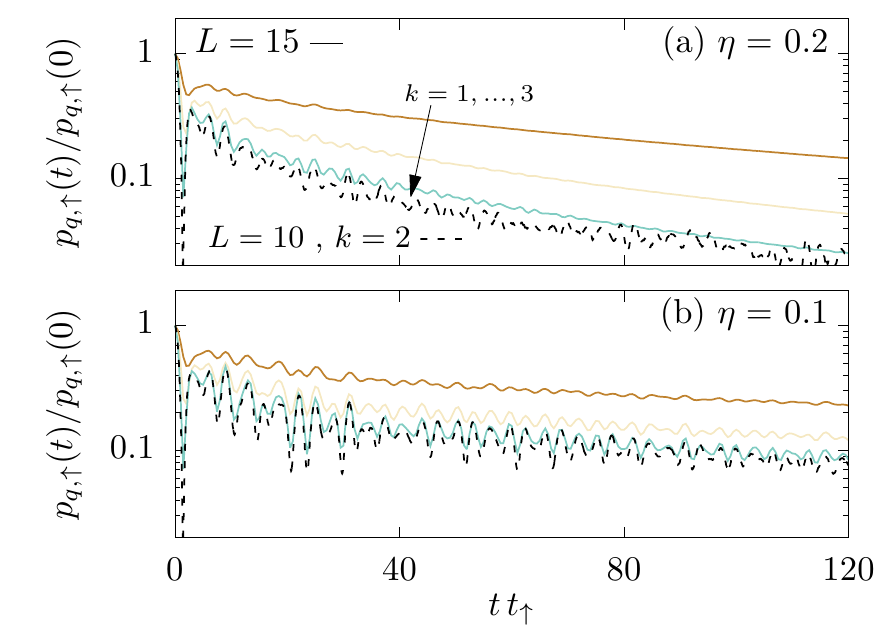}
	\caption{(Color online) Discrete Fourier transform $p_{q,\uparrow}(t)$ of the 
	density profile with momentum ${q=2\pi k/L}$ and wave numbers ${k=1,2,3}$ (arrow) 
	for two imbalance ratios (a) ${\eta=0.2}$ and (b) ${\eta=0.1}$ ($L=15$). Another 
	density mode $p_{q,\uparrow}(t)$ (dashed line) for a smaller system size ${L=10}$ 
	with wave number ${k=2}$ is shown for comparison, which has the same momentum ${q=2\pi/5}$ as 
	the mode ${k=3}$ for ${L=15}$.}
	\label{Fig8}
\end{figure}

Let us now turn to momentum-space dynamics again. To this end, Figs.\ \ref{Fig8}~(a) and (b) 
show the discrete Fourier modes $p_{q,\uparrow}(t)$ for imbalance ratios 
${\eta=0.2}$ and ${\eta=0.1}$. Note that the data is obtained for 
an even larger system with ${L=15}$ lattice sites and for momenta ${q=2\pi k/L}$ 
with ${k=1,2,3}$. 
Compared to \figref{Fig8b}, 
we find that the $p_{q,\uparrow}(t)$ now decay visibly slower 
for all wave numbers $k$. Moreover, in contrast to the scaling of decay rates in the case of 
normal diffusion [cf. Eq.~\eqref{eq:exp-decay-fourier}], the density modes now seem to decay at a 
similar rate for all $k$. Furthermore, even for small ${\eta=0.1}$ 
and ${k = 1}$, we find that $p_{q,\uparrow}(t)$ is clearly non-constant, which 
suggests that genuine localization is absent for ${\eta > 0}$.

To analyze the dependence on system size, \figref{Fig8} also 
shows the Fourier mode $p_{q,\uparrow}(t)$ for ${L=10}$ and wave 
number ${k=2}$.  
This mode has the same momentum ${q=2\pi/5}$ as the mode ${k=3}$ for ${L=15}$. 
We find that for both ${\eta=0.2}$ and ${\eta=0.1}$ the decay of 
$p_{q,\uparrow}(t)$ is almost independent of $L$. Especially for ${\eta = 
0.1}$, the curves show no significant differences
up to the maximum time ${\tth=120}$ shown.

\begin{figure}[tb]
	\centering
		\includegraphics[width=.45\textwidth]{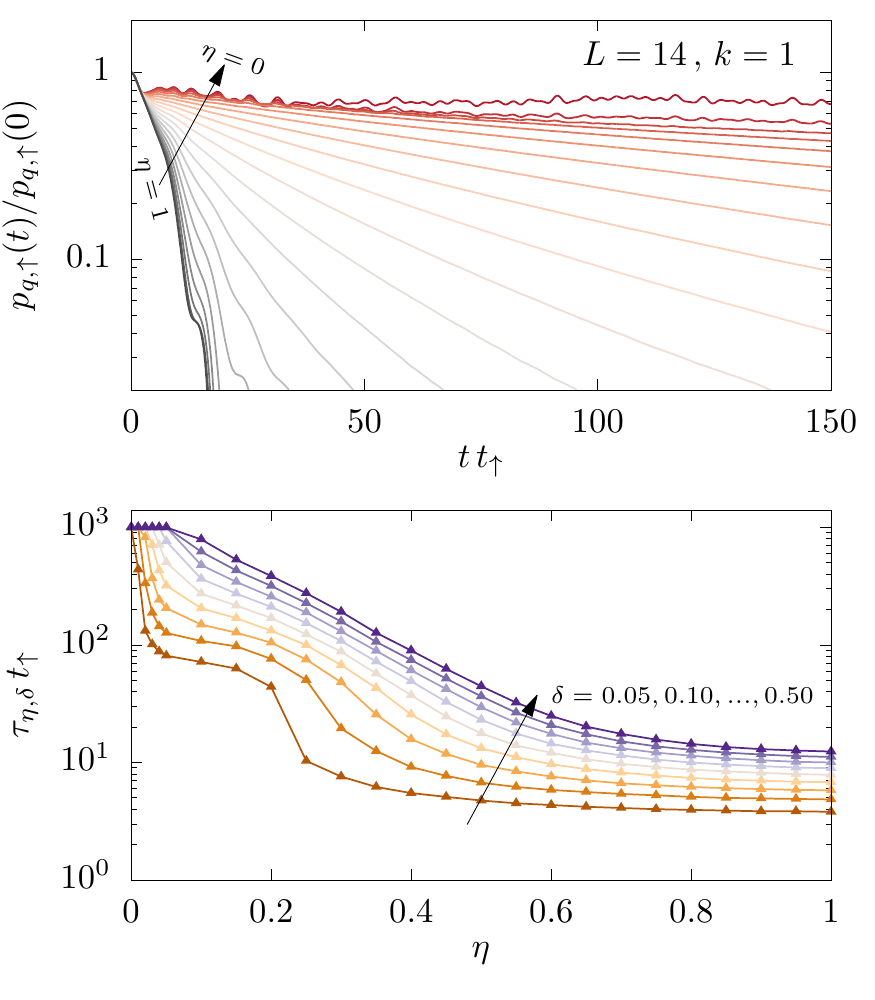}
	\caption{(Color online) (a) Discrete Fourier transform $p_{q,\uparrow}(t)$ of the density profile for fixed momentum ${q=2\pi/14}$ and
	(b) ``lifetime'' according to the definition \eqref{eq:Anderson-lifetime-av} for various distances ${\delta=0.05,\dots,0.5}$ in steps of $0.05$.}
	\label{Fig9}
\end{figure}

Finally, \figref{Fig9}~(a) shows the  
relaxation of the Fourier mode
$p_{q,\uparrow}(t)$ with the smallest wave number ${k=1}$ for various ${0 \leq \eta \leq 1}$.
The decay appears to be exponential for ${\eta\gtrsim0.2}$, 
albeit very slow for strong imbalances. 
While for sufficiently small times, all ${\eta > 0}$ curves agree with the ${\eta = 0}$ curve, 
they start to deviate at a certain point in time. 
In order to analyze this separation time from the 
${\eta=0}$ curve in more detail, 
we define
\begin{align}\label{eq:Anderson-lifetime-av}
	\tau_{\eta,\delta}=\max \left\lbrace \ t\ \vert\  \frac{\vert \overline{p}_{q,\uparrow}^{\eta}(t)-\overline{p}_{q,\uparrow}^0(t)\vert}{\overline{p}_{q,\uparrow}^0(t)} < \delta\ \right\rbrace
\end{align}
using the running averages of the density modes
\begin{align}
	\overline{p}_{q,\uparrow}^{\eta}(t)=\frac{1}{t}\int\limits_0^t p_{q,\uparrow}^{\eta}(t')dt'\ .
\end{align} 
It measures the maximum time up to which ${\eta=0}$ and ${\eta>0}$ curves do not deviate 
up to a distance $\delta$. (Note that this maximum time can not exceed the maximum simulation 
time, here ${t_{\max}\,t_{\uparrow}=1000}$. Moreover, the 
running averages are used to mitigate the fluctuations of the 
$p_{q,\uparrow}(t)$, which complicate the extraction of precise 
separation times.)

The physical picture for this analysis can be understood as 
follows. For very small but nonzero $\eta$, the heavy 
particles still appear as a quasi-static 
disorder potential for the lighter particles, which induces localization 
analogous to ${\eta = 0}$. At some point in time, however, the residual hopping 
of the heavy particles becomes relevant, which can be seen as an 
$\eta$-dependent ``lifetime'' 
of the Anderson insulator.
The corresponding data for different distances $\delta$ is shown in 
\figref{Fig9}~(b). For every $\delta$, the lifetime grows fast with 
decreasing $\eta$, but apparently is always finite for all $\eta$ 
considered. A complementary analysis of $\tau_{\eta,\delta}$, based on the spatial width $\Sigma(t)$ (cf. \figref{Fig6}), can be found in Appendix \ref{app:lifetime2} and provides a similar picture.

\section{Conclusion} \label{sec:Conclusion}

In this paper, we have studied the real-time dynamics of 
local charge densities in the Fermi-Hubbard chain with a 
mass-imbalance between the spin-$\uparrow$ and -$\downarrow$ particles. To this 
end, we have prepared a certain class of pure states featuring a sharp initial 
peak of the density profile for the (lighter) spin-$\uparrow$ particles in the 
middle of the chain and investigated the resulting non-equilibrium dynamics. 
Relying on dynamical quantum typicality, this dynamics can be 
related to time-dependent correlation functions at equilibrium. 

In the regime of weak and moderate imbalance, ${\eta\gtrsim0.6}$, we have provided evidence for the emergence of diffusive dynamics, manifesting in (i) Gaussian shape of density profiles, (ii) square-root scaling of the spatial variance in time, and (iii) exponentially decaying modes for small momenta. 

In contrast, in the regime of strong imbalance, ${\eta\lesssim0.6}$, we have 
observed signatures of anomalous transport, emerging as an exponential rather 
than a Gaussian shape of density profiles and subdiffusive scaling of spatial 
variance and density modes in time, consistent with other 
works \cite{Jin2015,Yao2016}. However, we cannot rule out 
that this anomalous transport is just a transient effect which crosses over to 
normal diffusion at even longer times, e.g., at time scales much longer than the ``lifetime'' of the Anderson insulator.

For very small but nonzero $\eta$, our results are consistent 
with the absence of genuine localization and support long but finite 
equilibration times. 

Promising future research directions include extensions of the model such as nearest-neighbor interactions and
the study of lower temperatures, including potential relations between static and dynamical properties at such temperatures \cite{Tezuka2012}.

\section*{Acknowledgments}

This work has been funded by the Deutsche Forschungsgemeinschaft (DFG) - Grants No.\ 397067869,  No.\ 355031190, and No.\ 397171440 - within the DFG Research Unit FOR 2692. 

\begin{appendix}

\section{Typicality relation}\label{app:typ-approx}

To make this paper 
self-contained, we here derive the typicality 
relation \eqref{eq:typicality-approximation}, see also \cite{Steinigeweg2017}. 
To this end, we start with the correlation function 
\begin{align}\label{eq:typ-approx-start}
	C_{r,\uparrow}(t)=2\braket{(n_{L/2,\uparrow}-\peq)(n_{r,\uparrow}(t)-\peq)}+\peq
\end{align}
and use ${\braket{n_{r,\uparrow}(t)}=\peq=1/2}$, while carrying out the multiplication of the brackets, to obtain
\begin{align}
	C_{r,\uparrow}(t)=2\braket{n_{L/2,\uparrow}n_{r,\uparrow}(t)}
	=\frac{\Tr{n_{L/2,\uparrow}n_{r,\uparrow}(t)}}{d\,/\,2}\ .
\end{align}
This expression, using cyclic invariance of the trace and the projection property ${n_{L/2,\uparrow}^2=n_{L/2,\uparrow}}$, can be written as
\begin{align}
	C_{r,\uparrow}(t)&=\frac{\Tr{n_{L/2,\uparrow}n_{r,\uparrow}(t)n_{L/2,\uparrow}}}{d\,/\,2}\ .
\end{align}
Exploiting typicality, the trace can be approximated by a single typical pure state $\ket{\phi}$ as  
\begin{align}
	C_{r,\uparrow}(t)&=\frac{\bra{\phi}n_{L/2,\uparrow}n_{r,\uparrow}(t)n_{L/2,\uparrow}\ket{\phi}}{\braket{\phi\,\vert\,\phi}/\,2}+\epsilon(\ket{\phi})\\\notag
	&\approx\frac{\left(\bra{\phi}n^{\dagger}_{L/2,\uparrow}\emath{\inice Ht}\right)n_{r,\uparrow}\left(\emath{-\inice Ht}n^{\phantom\dagger}_{L/2,\uparrow}\ket{\phi}\right)}{\braket{\phi\,\vert\,\phi}/\,2}\ ,
\end{align}
where the the variance of the statistical error $\epsilon(\ket{\phi})$ is bounded from above by ${\text{Var}(\epsilon)(\ket{\phi})<\mathcal{O}(1/d)}$  (at formally infinite temperature) and becomes negligibly small already for intermediate system sizes.
With ${\ket{\psi(0)}=n_{L/2,\uparrow}\ket{\phi}/\sqrt{\braket{\phi\,\vert\,\phi}/\,2}}$ we arrive at 
\begin{align}
	C_{r,\uparrow}(t)\approx\bra{\psi(t)}n_{r,\uparrow}\ket{\psi(t)}=p_{r,\uparrow}(t)
\end{align}
and finally, comparing to \eqref{eq:typ-approx-start}, 
\begin{align}
	p_{r,\uparrow}(t)-\peq\approx2\braket{(n_{L/2,\uparrow}-\peq)(n_{r,\uparrow}(t)-\peq)}\ .
\end{align}

\section{Equilibration for small $\eta$}\label{app:longtimes}
\begin{figure}[t]
	\centering
		\includegraphics[width=.45\textwidth]{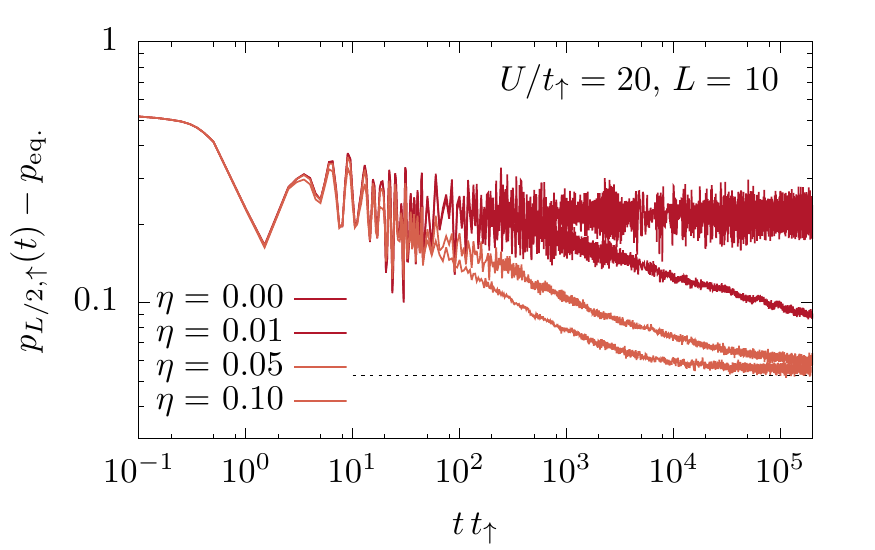}
	\caption{(Color online)  Decay of the central peak $p_{L/2,\uparrow}(t)$ for ${L=10}$ and long times at ${\eta\leq0.1}$.}
	\label{AppFig2}
\end{figure}

Complementary to \figref{Fig-center}, 
\figref{AppFig2} shows data for the central peak $p_{L/2,\uparrow}(t)$, but now 
for a smaller system size ${L=10}$ (in the half-filling sector 
$N_{\uparrow}+N_{\downarrow}=L$) and significantly longer time scales. 
We find that $p_{L/2,\uparrow}(t)$  ultimately decays towards its equilibrium 
value, even for very small values of $\eta$. Note that the interaction strength 
in \figref{AppFig2} is chosen as ${U/t_{\uparrow}=20}$, analogous to earlier 
investigations in Ref.~\cite{Yao2016}, where similar findings were presented 
for momentum-space observables.

\section{L-independence of density profiles}\label{app:L-independence}
\begin{figure}[b]
	\centering
		\includegraphics[width=.45\textwidth]{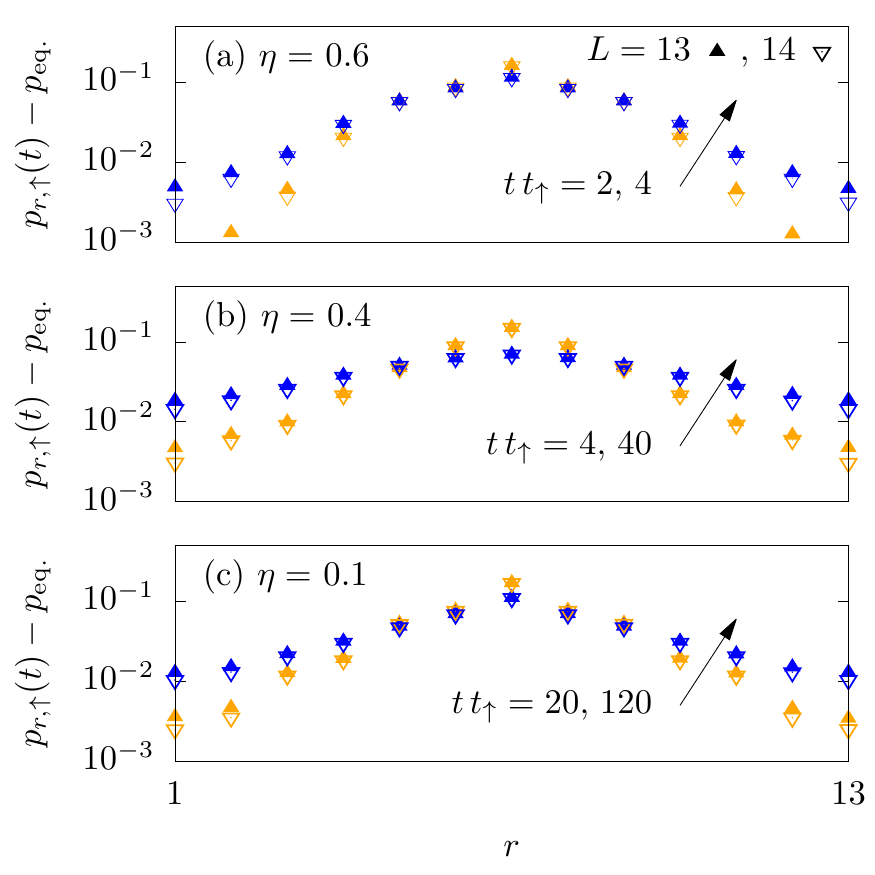}
	\caption{(Color online) Comparison of density profiles for two system sizes ${L=13,14}$ and a few exemplary imbalance ratios $\eta$. The overall behavior coincides nicely for both $L$, apart from slight deviations at the boundaries.}
	\label{AppFig1}
\end{figure}
To demonstrate the $L$-independence for the scaling of the density profiles,
\figref{AppFig1} shows $p_{r,\uparrow}(t)$ for two system sizes 
with ${L=13}$ and $14$ and exemplary  values for times $t$ and imbalances $\eta$.
Apart from small deviations at the tails, we find that the 
profiles for different $L$ are in very good agreement. This fact also 
demonstrates the accuracy of the typicality approach.

\section{Anderson lifetime}\label{app:lifetime2}
\begin{figure}[tb]
\centering
	\includegraphics[width=.45\textwidth]{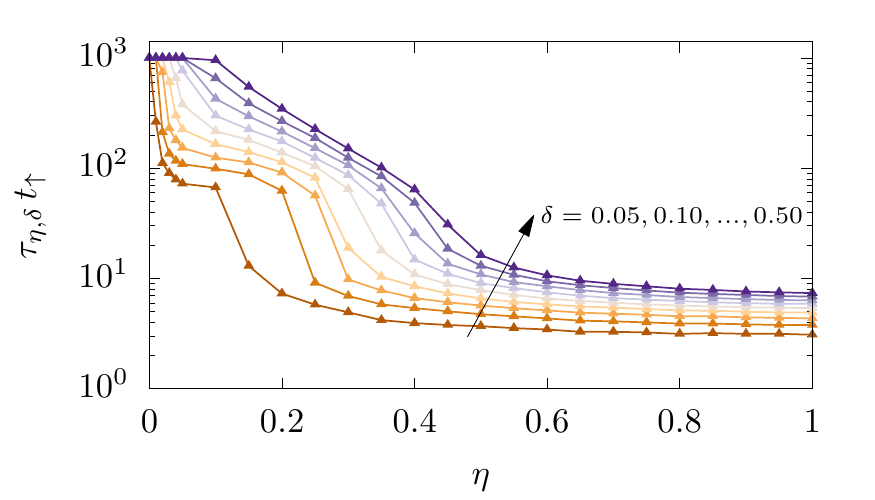}
\caption{(Color online) ``Lifetime'' in analogy to the definition \eqref{eq:Anderson-lifetime-av} for various distances ${\delta=0.05,\dots,0.5}$ in steps of $0.05$, now based on the spatial width $\Sigma(t)$ shown in \figref{Fig6}.}
\label{AppFig4}
\end{figure}
In addition to \figref{Fig9}~(b), 
\figref{AppFig4} shows another analysis of the lifetime $\tau_{\eta,\delta}$. Here, $\tau_{\eta,\delta}$ is calculated in analogy to \eqref{eq:Anderson-lifetime-av}, but based on the spatial width $\Sigma(t)$ (cf. \figref{Fig6}),
\begin{align}
	\tau_{\eta,\delta}=\max \left\lbrace \ t\ \vert\  \frac{\vert \overline{\Sigma}^{\eta}(t)-\overline{\Sigma}^0(t)\vert}{\overline{\Sigma}^{0}(t)} < \delta\ \right\rbrace
\end{align}
with
\begin{align}
	\overline{\Sigma}^{\eta}(t)=\frac{1}{t}\int\limits_0^t \Sigma^{\eta}(t')dt'\ .
\end{align} 
In comparison to \figref{Fig9}~(b), \figref{AppFig4} provides a very similar picture for the $\eta$-dependent lifetime.

\end{appendix}

\bibliographystyle{apsrev4-1}

\bibliography{paper}

\end{document}